\documentclass[letterpaper,aps,prd,preprint,nofootinbib,superscriptaddress]{revtex4}

\usepackage{color}

\usepackage[T1]{fontenc} 

\usepackage{graphicx,subfigure}
\usepackage{amsmath,amssymb}
\usepackage{bm}
\usepackage[utf8]{inputenc}

\usepackage{hyperref}
\hypersetup{ 
 setpagesize=false,
 bookmarksnumbered=true,
 colorlinks=true,
 linkcolor=blue,
 citecolor=red,
}

\newcommand{\bea}{\begin{eqnarray}}
\newcommand{\eea}{\end{eqnarray}}
\newcommand{\nn}{\nonumber}
\newcommand{\be}{\begin{equation}}
\newcommand{\ee}{\end{equation}}
\def\lsim{\raise0.3ex\hbox{$\;<$\kern-0.75em\raise-1.1ex\hbox{$\sim\;$}}}
\def\gsim{\raise0.3ex\hbox{$\;>$\kern-0.75em\raise-1.1ex\hbox{$\sim\;$}}}

\def\xHt{x_H}

\begin{document} 
\title{\boldmath  $R_K$ and $R_{K^*}$ in an aligned 2HDM with right-handed neutrinos }

\author{Luigi Delle Rose}
\email[]{L.Delle-Rose@soton.ac.uk}
\affiliation{\small INFN, Sezione di Firenze, and Dipartimento di Fisica ed Astronomia,	Universit\`a di  Firenze, Via G. Sansone 1, 50019 Sesto Fiorentino, Italy}
\affiliation{\small School of Physics and Astronomy, University of Southampton,
	Southampton, SO17 1BJ, United Kingdom}
	
\author{Shaaban Khalil}
\email[]{Skhalil@zewailcity.edu.eg}
\affiliation{\small Center for Fundamental Physics, Zewail City of Science and Technology, 6 October City, Giza 12588, Egypt}
	
\author{Simon J.D. King}
\email[]{sjd.king@soton.ac.uk}
\affiliation{\small School of Physics and Astronomy, University of Southampton,
	Southampton, SO17 1BJ, United Kingdom}
\affiliation{\small INFN, Sezione di Padova, and Dipartimento di Fisica ed Astronomia G. Galilei, Universit\`a di Padova, Via Marzolo 8, 35131 Padova, Italy}
\author{Stefano Moretti}
\email[]{s.moretti@soton.ac.uk}
\affiliation{\small School of Physics and Astronomy, University of Southampton,
	Southampton, SO17 1BJ, United Kingdom}
\affiliation{\small Particle Physics Department, Rutherford Appleton Laboratory, Chilton, Didcot, Oxon OX11 0QX, United Kingdom}

\date{\today}
\begin{abstract}
\footnotesize
We consider the possibility of explaining the recent $R_K$ and $R_{K ^*}$ anomalies in a two-Higgs doublet model, known as aligned, combined with a low-scale seesaw mechanism generating light neutrino masses and mixings. In this class of models, a large Yukawa coupling allows for significant nonuniversal leptonic  contributions, through box diagrams mediated by charged Higgs bosons and right-handed neutrinos, to the $b \to s \ell^+ \ell^-$ transition that can then account for both $R_K$ and $R_{K^*}$ anomalies.
\end{abstract}

\maketitle

\section{Introduction}
Recently, the LHCb Collaboration announced intriguing results \cite{Aaij:2019wad,Aaij:2017vbb} for the ratios $R_{K}= {\rm BR}(B^+\to K^+ \mu^+\mu^-)/{\rm BR}(B^+\to K^+ e^+ e^-)$
and $R_{K^*}= {\rm BR}(B^0\to K^{*0} \mu^+\mu^-)/{\rm BR}(B^0\to K^{*0} e^+ e^-)$. In fact, it was reported  that for two dilepton invariant mass-squared bins, $R_{K}$ and 
$R_{K^*}$ are given by
\bea
R_{K} &=& 0.846~ ^{+0.060 \, + 0.016}_{-0.054 \, -0.014} \quad \quad \textrm{for } 1.1 \textrm{ GeV}^2 \leq q^2 \leq 6 \textrm{ GeV}^2 \,, \nn \\
R_{K^*} &=& \left\{ \begin{array}{l}
0.66~ ^{+0.11}_{-0.07} ~\pm ~ 0.03 \quad \textrm{for } 0.045 \textrm{ GeV}^2 \leq q^2 \leq 1.1 \textrm{ GeV}^2 \\
0.69~ ^{+0.11}_{-0.07} ~\pm ~ 0.05 \quad \textrm{for } 1.1 \textrm{ GeV}^2 \leq q^2 \leq 6 \textrm{ GeV}^2
\end{array} \right.
\label{RKexp}
\eea
These measurements contradict the Standard Model (SM) expectations: $R_{K}^{\rm SM}\simeq R_{K^{*}}^{\rm SM} \simeq  1$ \cite{Bordone:2016gaq}, by a $\approx2.5 \sigma$ deviation. Hence, they are considered as important hints of new physics that violates lepton universality.  

\newcommand{\citeRK}{
\cite{Alonso:2014csa,
Hiller:2014yaa,Ghosh:2014awa,Glashow:2014iga,Hiller:2014ula,Gripaios:2014tna,Sahoo:2015wya,Crivellin:2015lwa,Crivellin:2015era,Celis:2015ara,Alonso:2015sja,Greljo:2015mma,Calibbi:2015kma,Falkowski:2015zwa,Carmona:2015ena,Chiang:2016qov,Becirevic:2016zri,Feruglio:2016gvd,Megias:2016bde,Becirevic:2016oho,Arnan:2016cpy,Sahoo:2016pet,Alonso:2016onw,Hiller:2016kry,Galon:2016bka,Crivellin:2016ejn,GarciaGarcia:2016nvr,Cox:2016epl,Jager:2017gal,Megias:2017ove,Crivellin:2017zlb,
Celis:2017doq,Becirevic:2017jtw,Cai:2017wry,Kamenik:2017tnu,Sala:2017ihs,DiChiara:2017cjq,Ghosh:2017ber,Alonso:2017bff,Greljo:2017vvb,
Bonilla:2017lsq,Feruglio:2017rjo,Ellis:2017nrp,Bishara:2017pje,Alonso:2017uky,Tang:2017gkz,Datta:2017ezo,Das:2017kfo,Dinh:2017smk,Bardhan:2017xcc,DiLuzio:2017chi,Chiang:2017hlj,Chauhan:2017ndd,King:2017anf,Chivukula:2017qsi,Dorsner:2017ufx,Buttazzo:2017ixm,Choudhury:2017qyt,Cline:2017ihf,Crivellin:2017dsk,
Guo:2017gxp,Chen:2017usq,Baek:2017sew,Bian:2017rpg,Megias:2017vdg,
Lee:2017fin,Assad:2017iib,DiLuzio:2017vat,
Calibbi:2017qbu,Cline:2017aed,Romao:2017qnu,Descotes-Genon:2017ptp,Altmannshofer:2017bsz,Bian:2017xzg,Cline:2017qqu,Botella:2017caf,DiLuzio:2017fdq,Barbieri:2017tuq,Sannino:2017utc,DAmbrosio:2017wis,Raby:2017igl,Blanke:2018sro,Falkowski:2018dsl,Arcadi:2018tly,CarcamoHernandez:2018aon,Marzocca:2018wcf,
Camargo-Molina:2018cwu,Bordone:2018nbg,Earl:2018snx,Matsuzaki:2018jui,
Becirevic:2018afm,Baek:2018aru,King:2018fcg,Kumar:2018kmr,Hati:2018fzc,Guadagnoli:2018ojc,deMedeirosVarzielas:2018bcy,Li:2018rax,Alonso:2018bcg,Azatov:2018kzb,DiLuzio:2018zxy,Duan:2018akc,Heeck:2018ntp,Angelescu:2018tyl,
Grinstein:2018fgb,Singirala:2018mio,Balaji:2018zna,Rocha-Moran:2018jzu,Kamada:2018kmi,Fornal:2018dqn,Geng:2018xzd,Kumar:2019qbv,
Baek:2019qte,Marzo:2019ldg,Cerdeno:2019vpd,Bhattacharya:2019eji,Ko:2019tts}}

Any signal of possible lepton nonuniversality would be striking evidence for physics beyond the SM (BSM). Therefore, the $R_{K}$ and 
$R_{K^*}$ anomalies attracted the attention of theoretical particle physicists and several new physics scenarios have been proposed to accommodate these results, see \citeRK.

A nontrivial flavor structure in the lepton sector is already required, beyond the SM, in order to explain the observation of neutrino oscillations.
Therefore, it is plausible and very attractive if the mechanism behind the $B$ anomalies can be related to the same physics responsible for the nonzero neutrino masses and oscillations.
In view of this, one then ought to explore extensions of the SM able to address both phenomena.

One of the possible scenarios for generating the observed lepton nonuniversality is to allow for large Yukawa couplings of the right-handed neutrinos with Higgs fields and charged leptons, as in a  
low-scale seesaw mechanism, with the inverse seesaw being one of the most notorious examples, for generating light neutrino masses. In this case, a nontrivial neutrino Yukawa matrix may lead to different results for BR$(B \to K e^+ e^-)$ and BR$(B \to K \mu^+ \mu^-)$. This framework has been recently considered in the supersymmetric (SUSY)  $B-L$ extension of the SM, where it was shown that that the box diagram mediated by a right-handed sneutrino, Higgsino-like chargino and light stop can account simultaneously for both $R_{K}$ and $R_{K^*}$ \cite{Khalil:2017mvb}.  In non-SUSY models, a similar box diagram can be obtained through a charged Higgs, instead of a chargino, and a right-handed neutrino, instead of a right-handed sneutrino. Therefore, a rather minimal model that can account for these discrepancies is an extension of the SM with two-Higgs doublets (so that we can have a physical charged Higgs boson) and right-handed neutrinos along with a low-scale seesaw mechanism (to guarantee large neutrino Yukawa couplings).   For earlier works in the context of two-Higgs Doublet models with or without extra neutral leptons see, for instance, Refs. \cite{Arnan:2017lxi,Campos:2017dgc,Iguro:2018qzf,Li:2018rax,Crivellin:2019dun}.
Note that there have been several attempts at explaining the above results through the penguin diagram as well, with a nonuniversal $Z'$ (see the previous list for examples, \citeRK) and also through tree-level mediation of a flavor-violating $Z'$ or leptoquark that induces a nonuniversal $b \to s \ell^+ \ell^-$ transition. 

The two-Higgs Doublet Model (2HDM), which is motivated by SUSY and grand unified theories (GUTs), is the simplest model that includes charged Higgs bosons. According to the types of couplings of the two-Higgs doublets to the SM fermions doublets and singlets, we may have a different type of 2HDMs. For example, if only one Higgs doublet couples to the SM fermions one obtains the type I 2HDM, 
while, in the case of one Higgs doublet coupling to the up quarks and the second Higgs doublet coupling to the down quarks and charged leptons, one obtains the type II 2HDM. 
Also, we may have a type III or IV 2HDM if both Higgs doublets couple to both up and down quarks as well as charged leptons.  
However, severe constraints are imposed on 2HDMs due to their large contributions to flavour changing neutral currents (FCNCs) that contradict the current experimental limits. Therefore, assumptions on the Yukawa couplings generated by different Higgs doublets are usually imposed. One of these assumptions is the alignment between the Yukawa couplings generated by $\Phi_1$ and $\Phi_2$, the two Higgs doublet fields. This class of models is called the aligned 2HDM (A2HDM). It is worth mentioning that, as discussed below, other 2HDMs cannot account for the LHCb results of lepton nonuniversality. 

In this paper, we emphasize that in the A2HDM, extended by (heavy) right-handed neutrinos in order to generate the (light) neutrino masses through a seesaw mechanism,  interesting results can be obtained for several flavor observables, such as $\mu \to e \gamma$, $B_s \to \mu \mu$ and, indeed,  $R_{K^{(*)}}$. In particular, one can account simultaneously for both aforementioned results on $R_K$ and $R_{K^*}$, through a box diagram mediated by a right-handed neutrino, top quark and  charged Higgs boson.

The paper is organized as follows. In Sec. II we introduce the main features of our A2HDM focusing on the neutrino sector and its interplay with the Higgs structures. 
(Here, we also briefly review some particular realizations of low-scale seesaw models for generating light neutrino masses.)  In Sec. III we discuss the most relevant constraints from flavor physics that affect our A2HDM parameter space. The calculation of the A2HDM contributions to $b \to s \ell^+ \ell^-$ transitions mediated by charged Higgs bosons and right-handed neutrinos is given in Sec. IV. Our numerical results are presented in Sec. V. Finally, our conclusions and remarks are given in Sec. VI.

\section{A2HDM}

The 2HDM is characterized by two Higgs doublets with hypercharge $Y=1/2$ which, in the Higgs basis, can be parameterized as 
\bea
\Phi_1 = \left( \begin{array}{c} G^+ \\ \frac{1}{\sqrt{2}}(v + \phi_1^0 + i G^0) \end{array} \right), \qquad
\Phi_2 = \left( \begin{array}{c} H^+ \\ \frac{1}{\sqrt{2}}(\phi_2^0 + i \phi_3^0) \end{array} \right),
\eea
where $v$ is the electroweak (EW) vacuum expectation value (VEV) and $G^+$ and $G^0$ denote the Goldstone bosons. The two doublets describe five physical scalar degrees of freedom which are given by the two components of the charged Higgs $H^\pm$ and three neutral states $\varphi_i^0 = \{h,H,A\}$, the latter obtained from the rotation of the $\phi_i^0$ fields into the mass eigenstate basis. The scalar squared mass matrix $\mathcal M_S^2$ is determined by the structure of the 2HDM scalar potential, see, for instance, Refs. \cite{Davidson:2005cw,Haber:2006ue,Haber:2010bw}, and diagonalized by the orthogonal matrix $\mathcal R$, where
\bea
\mathcal R \, \mathcal M_S^2 \, \mathcal R^T = \textrm{diag}(M_h^2, M_H^2, M_A^2) \,, \qquad \qquad  \varphi_i^0 = \mathcal R_{ij} \, \phi_i^0 \,.
\eea
In general, the three mass eigenstates $\varphi_i^0$ do not have definite $CP$ transformation properties, but in the $CP$-conserving scenario, $\phi_3^0$ does not mix with the other two neutral states and the scalar spectrum consists of a $CP$-odd field $A = \phi_3^0$ and two $CP$-even fields $h$ and $H$ that are defined from the interaction eigenstates through the two-dimensional orthogonal matrix
\bea
\left( \begin{array}{c} h \\ H \end{array} \right) = \left( \begin{array}{cc} \cos \alpha & \sin \alpha \\ -\sin \alpha & \cos \alpha \end{array} \right) \, \left( \begin{array}{c} \phi_1^0 \\ \phi_2^0 \end{array} \right) \,.
\eea

The most general Yukawa Lagrangian of the 2HDM can be written in the Higgs basis as
\bea
\label{eq:yukL}
- \mathcal L_Y &=&   \bar Q_L' \left( Y'_{1d}  \Phi_1 + Y'_{2d}  \Phi_2 \right) d_R' + \bar Q_L' \left( Y'_{1u} \tilde \Phi_1 + Y'_{2u} \tilde \Phi_2 \right) u_R'   \nn \\
 &+& \bar L'_L \left( Y'_{1\ell}  \Phi_1 + Y'_{2\ell}  \Phi_2 \right) \ell_R'  + \bar L'_L \left( Y'_{1\nu}  \tilde \Phi_1 +  Y'_{2\nu} \tilde \Phi_2 \right) \nu_R'  + \textrm{h.c.}
\eea
where the quark $Q_L', u_R', d_R'$ and lepton $L_R', \ell_R', \nu_R'$ fields are defined in the weak interaction basis and we also included the couplings of the left-handed lepton doublets with the right-handed neutrinos. The $\Phi_{1,2}$ fields are the two Higgs doublets in the Higgs basis and, as customary, $\tilde \Phi_i = i \sigma^2 \Phi_i^*$.
The Yukawa couplings $Y_{1j}'$ and $Y_{2j}'$, with $j = u,d,\ell$, are $3\times 3$ complex matrices while $Y_{1\nu}'$ and $Y_{2\nu}'$ are $3 \times n_R$ matrices, with $n_R$ being the number of right-handed neutrinos. In general, the Yukawas $Y_1'$ and $Y_2'$ cannot be simultaneously diagonalized in flavor space, so while the quark and the charged-lepton $Y_1'$ can be recast into a diagonal form in the fermion mass eigenstate basis, namely, $Y_{1} = \sqrt{2}/v M$, with $M$ being the fermion mass matrix, $Y_2$ would remain nondiagonal and thus give rise to potentially dangerous tree-level FCNCs. 
This problem is usually solved by enforcing that only one of the two Higgs doublets couple to a given right-handed field. This requirement is satisfied by  implementing a discrete $Z_2$ symmetry acting on the Higgs and fermion fields. There are four nonequivalent choices: types I, II, III, and IV (as previously  intimated). Another general way to avoid tree-level FCNCs in the Higgs sector is to require the alignment, in flavor space, of the two Yukawa matrices that couple to the same right-handed fermion \cite{Pich:2009sp}, namely,
\bea
Y_{2,d} = \zeta_d \, Y_{1,d} \equiv \zeta_d \, Y_d \,, \qquad Y_{2,u} = \zeta_u^* \, Y_{1,u} \equiv \zeta_u^* \, Y_{u}  \,, \qquad Y_{2,\ell} = \zeta_\ell \, Y_{1,\ell} \equiv \zeta_\ell \, Y_{\ell} \,,
\eea
where the proportionality constants $\zeta_f$ are arbitrary family universal complex parameters. This scenario is dubbed A2HDM.
The allowed sources of FCNCs at quantum level are highly constrained and the resulting structures are functions of the mass matrices and
Cabibbo-Kobayashi-Maskawa (CKM) matrix elements, so this model provides an explicit example of the popular minimal flavour violation (MFV) scenario \cite{DAmbrosio:2002vsn}.

Even though the alignment of the Yukawa matrices is strictly required, from observations, only in the quark and charged lepton sectors, 
we assume that the same mechanism which guarantees the aligned structure in the SM flavor space also holds  in the neutrino sector and leads to 
\bea
Y_{2,\nu} = \zeta_\nu^* \, Y_{1,\nu} \equiv \zeta_\nu^* \, Y_{\nu} \,.
\eea

In all sectors, the alignment is fixed to be exact at some specified scale $\mu _0$ and subsequently will misalign due to radiative corrections, as discussed in Refs. \cite{Jung:2010ik,Li:2014fea}. However, the flavor structure of the model constrains the nature of the new sources of FCNCs induced by renormalization group effects. Quantitatively, in the quark sector, these FCNC contributions are suppressed by mass hierarchies $m_q m_{q'}^2/v^3$ and provide negligible effects \cite{Jung:2010ik,Li:2014fea}. We will not consider the impact of the misalignment in this work.

Interestingly, $\zeta_f$ can provide new sources of $CP$ violation but in this work we will consider only real values. 
Notice also that the usual 2HDMs in which  tree-level FCNCs are removed by exploiting the discussed $Z_2$ discrete symmetry, namely, the types I, II, III, and IV, can be recovered for particular values of the proportionality constants $\zeta_f$ as shown in Table.~\ref{tab:2hdms}. 
\begin{table}[h]
\centering
\begin{tabular}{|ccccc|}
\hline
Aligned & Type I & Type II & Type III & Type IV \\
\hline \hline
$\zeta_u$ & $\cot \beta$ & $\cot \beta$ & $\cot \beta$ & $\cot \beta$ \\
$\zeta_d$ & $\cot \beta$ & $- \tan \beta$ & $-\tan \beta$ & $\cot \beta$ \\
$\zeta_l$ & $\cot \beta$ & $- \tan \beta$ & $\cot \beta$ & $-\tan \beta$ \\
\hline
\end{tabular}
\caption{Relation between the $\zeta_f$ couplings of the A2HDM and the ones of the $Z_2$ symmetric scenarios. \label{tab:2hdms}}
\end{table}

The Yukawa Lagrangian in Eq.~(\ref{eq:yukL}) generates a Dirac mass matrix for the standard neutrinos and can also be supplemented by a Majorana mass term $M_R'$ for the right-handed ones
\bea
- \mathcal L_{M_R} = \frac{1}{2} \nu_R'^{\, T} C M_R' \nu_R' + \textrm{h.c.}
\eea
where $C$ is the charge-conjugation operator. 
In particular, by exploiting a biunitary transformation in the charged-lepton sector and a unitary transformation on the right-handed neutrinos $L_L' = U_L \, L_L, \, \ell'_R = U_R^\ell \, \ell_R$ and $\nu'_R = U_R^\nu \, \nu_R$
is always possible to diagonalize (with real eigenvalues) the charged-lepton and Majorana mass matrices at the same time,
\bea
U_L^\dag Y'_{\ell} U_R^e &=& Y_{\ell} \equiv \frac{\sqrt{2}}{v} \textrm{diag} (m_e, m_\mu, m_\tau) \,, \nn \\
U_R^{^\nu \, T} M'_R U_R^\nu &=& M_R \equiv \textrm{diag}( M_1 , \ldots M_{n_R} ),
\eea
while $Y_{\nu} =  U_L^\dag Y'_{\nu} U_R^\nu$ remains nondiagonal. 
In this basis the neutrino mass matrix can be written as
\bea
\label{eq:mass_matrix}
- \mathcal L_{\mathcal M_\nu} = \frac{1}{2} N_L^T C \mathcal M N_L  + \textrm{h.c.}  = \frac{1}{2} (\nu_L^T \, \nu_R^{c \,\, T}) C  \left( \begin{array}{cc} 0 & M_D \\ M_D^T & M_R \end{array} \right)  \left( \begin{array}{c} \nu_L \\ \nu_R^c \end{array} \right)  \, ,
\eea
with $M_D = \frac{v}{\sqrt{2}} Y_{\nu}^*$ being the neutrino Dirac mass.
This can be diagonalized with the unitary $(3 + n_R) \times (3 + n_R)$ matrix $U$,
\bea
\left( \begin{array}{c} \nu_L \\ \nu_R^c \end{array} \right) = U \left( \begin{array}{c} \nu_l \\ \nu_h \end{array} \right) \equiv \left( \begin{array}{cc} U_{Ll} & U_{Lh} \\ U_{R^c l} & U_{R^c h} \end{array} \right) \left( \begin{array}{c} \nu_l \\ \nu_h \end{array} \right) ,
\eea
such that $\mathcal M_\nu = U^T \mathcal M U$ provides the masses of the three light active neutrinos $\nu_l$ and of the remaining $n_R$ heavy sterile neutrinos $\nu_h$.

The Yukawa interactions of the physical scalars with the mass eigenstate fermions are then described by
\bea
- \mathcal L_Y &=&  \frac{\sqrt{2}}{v} \bigg[   \bar u  ( - \zeta_u \, m_u \, V_{ud} \, P_L + \zeta_d \, V_{ud} \, m_d \, P_R  ) d   
+ \bar \nu_l  ( - \zeta_\nu \, m_{\nu_l} \, U^\dag_{L l}  \,  P_L   + \zeta_\ell  \,  U^\dag_{L l}  \, m_\ell \, P_R )  \ell       \nn \\
&+&  \bar \nu_h  ( - \zeta_\nu \, m_{\nu_h} \, U^\dag_{L h}  \,  P_L   + \zeta_\ell  \,  U^\dag_{L h}  \, m_\ell \, P_R )  \ell   \bigg]  H^+  + \textrm{h.c.}  \nn \\
&+& \frac{1}{v} \sum_i \sum_{f=u,d,\ell} \xi_f^i \, \varphi_i^0 \, \bar f  \, m_f \, P_R \,  f   \nn \\
&+& \frac{1}{v} \sum_i \xi_\nu^i \, \varphi_i^0 (\bar \nu_l \, U_{Ll}^\dag + \bar \nu_h \, U_{Lh}^\dag) P_R (U_{Ll} \, m_{\nu_l} \, \nu_l^c + U_{Lh} \, m_{\nu_h} \, \nu_h^c) + \textrm{h.c.} 
\eea
where the couplings of the neutral Higgs states to the fermions are given by 
\bea
\xi_{u, \nu}^i = \mathcal R_{i1} + ( \mathcal R_{i2} - i  \mathcal R_{i3} ) \zeta_u^*   \,, \qquad
\xi_{d,\ell}^i = \mathcal R_{i1} + ( \mathcal R_{i2} + i  \mathcal R_{i3} ) \zeta_{d,\ell} .
\eea
Because of the alignment of the Yukawa matrices all the couplings of the scalar fields to fermions are proportional to the corresponding mass matrices.
Finally, the weak neutral and charged interactions of the neutrinos are
\bea
 \mathcal L_Z &=& \frac{g}{2 \cos \theta_W}  (\bar \nu_l \, U_{Ll}^\dag + \bar \nu_h \, U_{Lh}^\dag) \gamma^\mu  (U_{Ll} \, \nu_l  +  U_{Lh} \, \nu_h )  Z_\mu, \nn \\
 \mathcal L_W &=& - \frac{g}{\sqrt{2}} \left[  (\bar \nu_l \, U^\dag_{L l} + \bar \nu_h \, U^\dag_{L h}) \gamma^\mu P_L \, \ell \right] W^{+}_\mu  + \textrm{h.c.}
\eea

In this paper, rather than presenting a complete model in the neutrino sector by specifying the structure and the hierarchies of the neutrino mass matrices, we work in a simplified framework that captures the interesting phenomenology while preserving a significant degree of model independence. 
In particular, we consider a single extra heavy neutrino despite the usual requirement of additional sterile states to fully accommodate the observed pattern of the light neutrino masses and mixing angles. 
Indeed, low-scale right-handed neutrinos with sizeable mixings with the SM left-handed neutrino states, such that they may provide visible effects in physical observables at the EW scale, usually affect the light neutrino masses with inadmissible large contributions. This issue is nicely solved in extended seesaw models \cite{Mohapatra:1986bd,GonzalezGarcia:1988rw,Pilaftsis:1991ug,Mohapatra:2005wg,Malinsky:2005bi,deGouvea:2006gz,Kersten:2007vk,Forero:2011pc,Dev:2012sg}. For example the linear or  inverse seesaw mechanism, in which extra sterile neutrino states are introduced to allow for large mixings while correctly reproducing the observed smallness of the light neutrino masses. \\
For the sake of definiteness, in the following section we briefly present some specific setup that could be employed to realize the phenomenological scenario described above.

\subsection{Some explicit examples of low-scale seesaw mechanism}
As stated above, in order to achieve a sizeable mixing with the heavy sterile neutrino while avoiding, at the same time, large contributions to the light neutrino masses, it is necessary 
to require the neutrino Majorana mass matrix and the Dirac neutrino Yukawa coupling to realize a particular structure. 
To match the nomenclature usually employed in the literature, we split the set of right-handed neutrinos defined above into two classes, one of $n_S$ SM-singlet fermionic fields $S^i$ and another one of fermionic states that we continue to call right-handed neutrinos. The differences between the two will be clear in a moment. 

In the basis $N_L = (\nu_L, \nu_R^c,S)^T$, the mass matrix can be parametrized as 
\bea
\mathcal M = \left( 
\begin{array}{ccc}
0 & m_D & m_S \\
m_D^T & m_N & m_R \\
m_S^T & m_R^T & \mu_S
\end{array} 
\right)
\eea
which has the same structure as the one given in Eq.~(\ref{eq:mass_matrix}) provided that 
\bea
M_D \equiv (m_D, m_S)\,, \qquad  M_R \equiv \left( \begin{array}{cc} m_N & m_R \\ m_R^T & \mu_S \end{array} \right) \,.
\eea
The $m_D$ and $m_S$ are, respectively, $3 \times n_R$ and $3 \times n_S$ mass matrices mediating the interactions between the charged leptons and the right-handed and sterile neutrinos while $m_N$, $m_R$ and $\mu_S$ are $n_R \times n_R$, $n_R \times n_S$ and $n_S \times n_S$ mass matrices, respectively. 
As both right-handed $\nu_R$ and sterile $S$ neutrinos can be assigned lepton number $L = 1$, 
the mass terms $m_N$, $m_S$ and $\mu_S$ violate lepton number by two units. 

Two commonly studied mass patterns are the inverse and  linear seesaws, which are characterized by $m_S = m_N = 0$ and $\mu_S = m_N = 0$, respectively.
In these cases, the vanishing of the Majorana mass $\mu_S$ or $m_S$ would restore lepton number conservation and, as such, would increase the symmetry of the model. This feature makes the two masses naturally small accordingly to the 't Hooft naturalness principle.

Following the standard seesaw calculation and by assuming the hierarchy $\mu_S \, (m_S) \ll m_D, m_R$ for the inverse (linear) seesaw scenario, the $3 \times 3$ light neutrino mass matrix is
\bea
\label{eq:mlight_iss} 
m_\textrm{light} \simeq  \left\{ \begin{array}{ll}
m_D (m_R^T)^{-1} \mu_S \, m_R^{-1} m_D^T  & \qquad \textrm{inverse seesaw} \\
m_S m_R^{-1} m_D^T + m_D (m_R^T)^{-1} m_S^T & \qquad \textrm{linear seesaw}
\end{array} \right.
\eea
which is diagonalized by the Pontecorvo-Maki-Nakagawa-Sakata (PMNS) matrix $U_\textrm{PMNS}$, namely 
\bea
U_\textrm{PMNS}^T m_\textrm{light} U_\textrm{PMNS} = m_\nu \equiv \textrm{diag}(m_{\nu_1}, m_{\nu_2}, m_{\nu_3}).
\eea 

Differently from the standard type I seesaw case, in which $m_\nu \sim m_D^2/m_R$ with $m_D \ll m_R$, the lightness of the active neutrino masses is ensured in these low-scale seesaw scenarios 
by the smallness of the $\mu_S$ $ (m_S)$ parameters. This feature prevents $m_D$ from being extremely suppressed with respect to the Majorana mass and, as such, may allow for non-negligible couplings between the heavy neutrinos and the SM gauge bosons which are set by the mixing $U_{Lh}$.  

To understand the dependence of the mixing $U_{Lh}$ of the left-handed SM neutrinos with the extra sterile states,
it is instructive to study the mixing matrix $U$ in the limit of negligible $\mu_S~(m_S)$ and small $m_D/m_R$. 
While the first requirement is necessary to reproduce the lightness of the active neutrino states, the latter is used here only to simplify the structure of $U$ which reads as
\bea
\label{eq:Umix}
U = \left( \begin{array}{ccc}
1 & \frac{1}{\sqrt{2}} m_D^* m_R^{-1} & \frac{i}{\sqrt{2}} m_D^* m_R^{-1} \\
0 & \frac{1}{\sqrt{2}} & - \frac{i}{\sqrt{2}} \\
-m_R^{-1} m_D^T & \frac{1}{\sqrt{2}} &  \frac{i}{\sqrt{2}} \\
\end{array} \right) + \mathcal O\left( \frac{m_D^2}{m_R^2}\right) \,.
\eea
Notice that the PMNS matrix has been set to the unit 1, consistently with the approximation $m_{\nu_i} \simeq 0$. 
From Eq.~(\ref{eq:Umix}), one can immediately realize that $U_{Lh}$ is set, as naively expected from dimensional arguments, by the ratio $m_D/m_R$.

Once some specific inputs are provided for $m_D$ and $m_R$, the corresponding $\mu_S~(m_S)$ matrix that ensures the agreement with the light neutrino mass splittings and mixing angles can always be reconstructed from Eq.~(\ref{eq:mlight_iss}). 
For instance, in the inverse seesaw case, we find
\bea
\mu_S = m_R^T m_D^{-1} U_\textrm{PMNS}^* m_\nu U_\textrm{PMNS}^\dag (m_D^{T})^{-1} m_R, 
\eea
where $m_\nu$ and $U_\textrm{PMNS}$ are chosen in agreement with the bounds from the low-energy neutrino data which we report below for the sake of completeness. 
In particular, one should enforce the following constraints from the latest results of the $\mathcal{V}$fit group \cite{Esteban:2016qun} extracted from the $\mathcal{V}$fit 3.2 (2018) data. \\ \\
1) {\bf Neutrino mass squared differences} \\ 
The $3\sigma$ confidence-level (CL)  ranges on the mass squared differences
\begin{align}
 \Delta m_{21}^2 &= (6.80 \to 8.02) \times 10^{-5} \, \textrm{eV}^2 \nn \\
 \Delta m_{3l}^2 &= \left\{ \begin{array}{l} (2.399 \to 2.593) \times 10^{-3} \, \textrm{eV}^2 \quad \textrm{(for $l=1$ N.O.)}  \\ (- 2.562 \to -2.369) \times 10^{-3} \, \textrm{eV}^2 \quad \textrm{(for $l=2$ I.O.)}   \end{array} \right.
\end{align}
where the first and second possibility refer to the assumption of normal (N.O.) and inverted ordering (I.O.) in the light neutrino masses, respectively.  \\
2) {\bf Leptonic mixing matrix} \\
The $3\sigma$ CL ranges on the elements of the leptonic mixing matrix $U_\textrm{PMNS}$
\begin{align}
\sin^2 \theta_{12} &= (0.272 \to 0.346) \nn \\
 \sin^2 \theta_{23} &= \left\{ \begin{array}{l} (0.418 \to 0.613) \quad \textrm{N.O.} \\ (0.435 \to 0.616) \quad \textrm{I.O.}  \end{array} \right. \quad 
 \sin^2 \theta_{13} &= \left\{ \begin{array}{l} (0.01981 \to 0.02436) \quad \textrm{N.O.} \\ (0.02006 \to 0.02452) \quad \textrm{I.O.}  \end{array} \right. 
\end{align} \\ \\
Notice also that $m_D$ and $m_R$ cannot be chosen freely since the $U_{Lh}$ block of the mixing matrix that they define is constrained by the unitarity requirement
that directly affects the analysis presented in this work. The corresponding bound is discussed in the next section together with all the other relevant constraints.

\section{Relevant parameter space and constraints}

\subsection{Unitarity bounds on the neutrino mixing matrix}

The $3 \times 3$ block of the mixing matrix $U$ corresponds to a nonunitary $\tilde U_\textrm{PMNS}$ matrix. The bounds on the deviation from unitarity of $\tilde U_\textrm{PMNS}$ have been obtained in Refs. 
\cite{Antusch:2014woa,Antusch:2016brq} using an effective field theory approach in which the masses of the heavy neutrinos lie above the EW scale. This constraint can be recast as follows
\begin{align}
\label{eq:unitarity}
& \epsilon_{\alpha \beta} \equiv \left| \sum_{i}^{n_R} U^*_{\alpha i} U_{\beta i} \right| \equiv \left| \sum_{i = 1}^{n_R} (U^*_{Lh})_{\alpha i} (U_{Lh})_{\beta i} \right| = \left| \delta_{\alpha \beta} - (\tilde U_\textrm{PMNS}^\dag \tilde U_\textrm{PMNS} )_{\alpha \beta}\right|, \nn \\
 & \left| \tilde U_\textrm{PMNS} \tilde U_\textrm{PMNS}^\dag \right| = \left( \begin{array}{ccc}
(0.9979 \to 0.9998) & < 10^{-5} & < 0.0021  \\
< 10^{-5} & (0.9996 \to 1.0) & < 0.0008 \\
< 0.0021 & < 0.0008 & (0.9947 \to 1.0)
\end{array} \right) \,.
\end{align}

\subsection{Lepton flavor-violating processes}

We consider the lepton flavor-violating decays $\ell_\alpha \rightarrow \ell_\beta \gamma$ induced at one-loop order by the sterile neutrinos and check their compatibility with the experimental upper bounds at 90\% CL \cite{Tanabashi:2018oca},
\begin{equation}
\textrm{BR}(\mu \to e \gamma) \le 4.2 \times 10^{-13} \,, ~~~
\textrm{BR}(\tau \to e \gamma) \le 3.3 \times 10^{-8} \,, ~~~
\textrm{BR}(\tau \to \mu \gamma) \le 4.4 \times 10^{-8}\,.
\end{equation}
The branching ratios (BRs) of the aforementioned decay rates are given by
\bea
\hspace{-0.5cm}
 \textrm{BR}(\ell_\alpha \to \ell_\beta \gamma)  = 
\mathcal C \, \left| \sum_{i = 1}^{n_R} (U^*_{Lh})_{\alpha i} (U_{Lh})_{\beta i}  \left[ \mathcal G_{W^\pm} \left( \frac{m_{\nu_{h_i}}^2}{M_W^2} \right)   +   \mathcal G_{H^\pm} \left( \frac{m_{\nu_{h_i}}^2}{M_{H^\pm}^2} \right)  \right] \right|^2 
\label{eq:BRltolgamma}
\eea
with 
\bea
\mathcal C =  \frac{\alpha_W^3 s_W^2}{256 \pi^2} \left( \frac{m_{\ell_\alpha}}{M_W} \right)^4 \frac{m_{\ell_\alpha}}{\Gamma_{\ell_\alpha}} 
\eea
where $\Gamma_{\ell_\alpha}$ is the total decay width of the lepton $\ell_\alpha$ and the loop functions are
\bea
\mathcal G_{W^\pm}(x) &=& \frac{-x + 6 x^2 - 3 x^3 - 2 x^4 + 6 x^3 \log x}{4(x - 1)^4} \,, \nn \\
\mathcal G_{H^\pm}(x) &=& \frac{\zeta_\nu^2}{3} \mathcal G_{W^\pm}(x) + \zeta_\nu \zeta_\ell \frac{ x (-1 + x^2 - 2 x \log x) }{2 (x - 1)^3} \,,
\label{eq:GHpm}
\eea
where we have neglected the mass of the lepton in the final state (for a study of lepton flavor-violating processes in the context of low-scale seesaw models, see, for instance, Refs. \cite{Arganda:2014dta,Arganda:2015naa,Arganda:2015ija,DeRomeri:2016gum}). \\
The $\mathcal G_{H^\pm}$ can offer large contributions, larger than  $\mathcal G_{W^{\pm}}$, for sizeable values of the couplings $\zeta _\nu, \zeta _l$, which are, as such, strongly constrained by lepton flavor-violating processes. These can be tamed by controlling the size of the mixing matrix elements $(U_{Lh})_{\alpha i}$ which should 
 be highly suppressed in order to avoid any large contribution to these sensitive processes.

\subsection{Flavour constraints from meson processes}
Here we briefly mention the relevant constraints from measurement of flavor observables in meson mixing and decays. 
These have been studied in the context of general 2HDMs and the majority of the bounds extracted from these can be straightforwardly applied in our case since the presence of the sterile neutrinos does not add any significant contribution at leading order. In particular, the 2HDM with the alignment in the flavor sector has been scrutinized in \cite{Jung:2010ik,Jung:2010ab,Li:2014fea,Enomoto:2015wbn} to which we refer for delineating the allowed parameter space spanned by the $\zeta_{u,d,\ell}$ couplings and the charged Higgs mass $M_{H^\pm}$. 
\begin{itemize}
\item {\bf Neutral meson mixing} \\
{
The meson mixing observables $\Delta M_s$, $\Delta M_d$ and $|\epsilon_K|$ constrain large values of the $\zeta_u$ parameter and smaller values of $m_{H ^\pm}$, due to the box diagram with two charged Higgses, or $H^\pm W ^\mp$. In turn, they do not significantly affect $\zeta_d$, which dependence is suppressed by $m_b^2/M_W^2$ in the first two observables, while does not appear at all in the third one. 
Explicit formulas for the corrections to the SM predictions of these observables in the 2HDM are lengthy and can be found in Ref. \cite{Enomoto:2015wbn}. 
Here, we present only the results for $\Delta M_q$ since the $B_q^0 - \bar B_q^0$ mixing is the one mostly affecting our parameter space,
\bea
\Delta M_q = \frac{G_F^2}{24\pi^2} M_W^2 M_{B_q} |V_{tq} V_{tb}^*|^2 f_{B_q}^2 \left[   \hat B_{B_q} \, \eta_{B_q} \, C_V +    \hat B_{B_q}^{ST} \, \eta_{B_q}^{ST} \, C_{ST}     \right],
\eea
where $\hat B_{B_q}$ and $\hat B_{B_q}^{ST}$ are the bag parameters while $\eta_{B_q}$ and $\eta_{B_q}^{ST}$ account for the running of the QCD corrections. Their values can be found in Ref. \cite{Enomoto:2015wbn}.
The Wilson coefficients $C_V$ and $C_{ST}$ arise, respectively, from the vector operator and from a combination of the scalar and tensor operators. Their  leading contributions can be parameterised as
\bea
C_V &=& x_t \left[ A_{WW} + 2 x_t A_{WH} + x_t A_{HH} \right] \,, \nn \\
C_{ST} &=&4  x_b  x_t^2 \left[ A_{WH}^{ST} + A_{HH}^{ST} \right],
\eea
with $x_q = M_q^2/M_{W}^2$. The term $A_{WW}$ originates from the SM while all the others are induced by the exchange of one or two charged Higgses in the box diagrams.
The SM also contributes to the Wilson coefficient $C_{ST}$ but suffers from a large suppression by the bottom quark mass. In contrast, this suppression can be alleviated in the 2HDM by large factors of $\zeta_{u,d}$.
In our scenario, the main corrections to the SM prediction appear in $C_V$ and, in particular, in $A_{WH} $ since $A_{HH}$ is suppressed by an additional $H^+$ propagator in the box diagram. The relevant coefficients are
\bea
A_{WW} &=& 1 + \frac{9}{1 - x_t} - \frac{6}{(1 - x_t)^2} - 6 x_t^2 \frac{\log x_t}{(1 - x_t)^3} \,, \nn \\
A_{WH} &=& \zeta_u^2 \left[  \frac{x_H \left(x_t-4\right)}{x_t \left(x_H-1\right) \left(x_t-1\right) }      +      \frac{x_H \left(x_t-4 x_H\right) \log x_H}{x_t \left(x_H-1\right){}^2 
   \left(x_H-x_t\right)}+\frac{3 x_H \log x_t}{\left(x_t-1\right){}^2
   \left(x_H-x_t\right)}      \right] \,, \nn \\
A_{HH} &=& \zeta_u^4 \left[ \frac{x_H (x_H + 1)}{x_t (x_H - 1)^2 } - \frac{2 x_H^2 \log x_H }{x_t (x_H - 1)^3 }  \right] \,.
\eea
We require that our benchmark scenario, detailed below, maximizes the impact on $R_K$ and $R_{K^*}$ while complying with the observed $B_q ^0 - \bar{B} _q ^0$ mixing at the level of $2 \sigma$. 
A lighter charged Higgs boson or a larger $\zeta_u$ would spoil the constraint from $\Delta M_s$.
}
Obviously, there is no dependence on $\zeta_\ell$ and $\zeta_\nu$.
\item {\bf Radiative $B_s \to X_s \gamma$ decay} \\
The $B_s \to X_s \gamma$ decay rate represents one of the best measured observables and it is employed to constrain several new physics scenarios.
The contribution of the charged Higgs boson is encoded in the Wilson coefficients $C_7$ and $C_8$. 
At leading order, the corresponding new physics corrections are sensitive to $\zeta_u$ and $\zeta_d$ and are given by
\begin{align}
C_i = \frac{(\zeta_u)^2}{3} G_i^1 \left(\frac{M_t^2}{M_{H^+}^2} \right) + \zeta_u \zeta_d \, G_i^2 \left(\frac{M_t^2}{M_{H^\pm}^2} \right) 
\end{align}
with
\begin{align}
G_7^1(x) &= \frac{y(7-5y-8y^2)}{24(y-1)^3} + \frac{y^2(3y-2)}{4(y-1)^4} \log x , \quad G_7^2(x) = \frac{y(3-5y)}{12(y-1)^2} + \frac{y(3y-2)}{6(y-1)^3} \log x \,, \nonumber \\
G_8^1(x) &= \frac{y(2+5y-y^2)}{8(y-1)^3} - \frac{3y^2}{4(y-1)^4} \log x , \quad G_8^2(x) = \frac{y(3-y)}{4(y-1)^2} - \frac{y}{2(y-1)^3} \log x \,.
\end{align}
As for the new physics contributions to the meson mixing observables, the values of the $\zeta_\ell$ and $\zeta_\nu$ parameters are completely irrelevant in the determination of the $b \to s \gamma$ transition.
\item {\bf Leptonic decay of the neutral mesons $B^0_q \to \mu^+ \mu^-$} \\
The BR of the $B^0_q \to \mu^+ \mu^-$ meson decay is given by
\bea
 \textrm{BR}(B^0_q \to \mu^+ \mu^-) = \textrm{BR}_\textrm{SM}(B^0_q \to \mu^+ \mu^-) (|P|^2 + |S|^2) 
\eea
with
\bea
\hspace{-0.75cm}
 P = \frac{C_{10}}{C_{10}^\textrm{SM}} + \frac{m_{B^0_q}^2}{2 M_W^2} \left( \frac{m_b}{m_b + m_q} \right) \frac{C_P}{C_{10}^\textrm{SM}} \,, ~
 S = \sqrt{1 - \frac{4 m_\mu^2}{m_{B^0_q}^2}} \frac{m_{B^0_q}^2}{2 M_W^2} \left( \frac{m_b}{m_b + m_q} \right) \frac{C_S}{C_{10}^\textrm{SM}} \,.
\eea
The leading new physics contribution appears in the Wilson coefficient $C_{10}$ and it is mediated by the charged Higgs in 2HDMs. Interestingly, the same coefficient is also corrected by the presence of the heavy sterile neutrinos and it is sensitive to, besides $\zeta_{u,d,\ell}$, the coupling $\zeta_\nu$ of the charged Higgs to the sterile neutrino states.
The impact of the $C_P$ and $C_S$ coefficients is, in contrast, suppressed by the $m_{B^0_q}^2/M_W^2$ factor unless the former can be enhanced with respect to $C_{10}^\textrm{SM}$ as, for instance, in a $Z_2$ symmetric model with large $\tan \beta$. In the flavor aligned 2HDM and in the parameter space in which we are interested, namely, $\zeta_u \simeq 1$ and $\zeta_d \simeq \zeta_\ell \simeq 0$, the bound from the measurement of the $B^0_s \to \mu^+ \mu^-$ transition is usually weaker than the one from the $B_s \to X_s \gamma$ decay. Nevertheless, due to the contribution from the heavy sterile neutrinos which is proportional to the new parameter $\zeta_\nu$, we recompute the corresponding constraint by employing the \texttt{FLAVIO} package \cite{Straub:2018kue}.
\item {\bf Leptonic decay of the charged mesons $M^\pm \to \tau^\pm \nu$} \\
The $M^\pm \to \tau^\pm \nu$ decay occurs at tree level through charged current processes and the corresponding BR is
\bea
 \textrm{BR}(M^\pm \to \tau^\pm \nu) = \frac{\tau_M \, G_F^2 \, m_M \, m_\tau^2}{8 \pi} \left( 1 - \frac{m_\tau^2}{m_M^2} \right)^2 |V_{ud}|^2 f_M^2 |1 + C_H|^2,
\eea
where $f_M$ and $\tau_M$ are the decay constant and the lifetime, respectively, and the contribution of the charged Higgs boson is encoded in
\bea
C_H = \frac{\zeta_u \, \zeta_\ell \, m_u - \zeta_d \, \zeta_\ell \, m_d}{m_u + m_d} \frac{m_M^2}{M_{H^\pm}^2} \,.
\eea
The most constraining decay mode is found to be $B \to \tau \nu$ which is, however, only relevant for light charged Higgs masses.
\end{itemize}

\section{Contributions to  $b \to s \ell^+\ell^-$ processes}
The effective Hamiltonian for the $b \to s \ell^+\ell^-$ transitions is given by
\bea
\mathcal H_\textrm{eff} = - \frac{4 G_F}{\sqrt{2}} \frac{\alpha}{4 \pi} V_{ts}^* V_{tb} \sum_{i} C_i \mathcal O_i + \textrm{h.c.}
\eea
where the relevant operators for the analysis of the $R_K$ and $R_{K^*}$ anomalies are 
\bea
\mathcal O_9 = (\bar b \gamma^\mu P_L s) (\bar \ell \gamma_\mu \ell) \,, \qquad  \mathcal O_{10} = (\bar b \gamma^\mu P_L s) (\bar \ell \gamma_\mu \gamma_5 \ell) \,.
\eea
The new physics effects in the corresponding Wilson coefficients can be recast as
\begin{align}
\Delta C_i &= C_i^{(\Delta_Z)}(H^\pm) + C_i^{(\Delta_Z)}(N_R)  		& Z - \textrm{penguin}		\nn \\
&+ C_{i}^{(\Delta_\gamma)}(H^\pm)  + C_{i}^{(\Delta_\gamma)}(N_R)	& \gamma - \textrm{penguin} 		 \nn \\
&+ C_{i}^{(\Box)}(H^\pm) + C_{i}^{(\Box)}(N_R) + C_{i}^{(\Box)}(N_R, H^\pm)  & \textrm{box}  \,
\end{align}
where $\Delta_Z$, $\Delta_\gamma$ and $\Box$ denote the contributions from the $Z$ penguins, the photon penguins and the box diagrams, respectively. 
Moreover, $C_i^{(\bullet)}(H^\pm)$ represents the charged Higgs contribution typical of the 2HDM, $C_i^{(\bullet)}(N_R)$ denotes the loop corrections from heavy sterile neutrinos and $W^\pm$ bosons that is present in the seesaw extensions of the SM without extra Higgses, and $C_i^{(\bullet)}(N_R,H^\pm)$ represents a combined contribution from diagrams with both sterile neutrinos and charged Higgs.  \\
As the sterile neutrinos are not charged under the (color) $SU(3)$ gauge group, their contribution to the penguin diagrams is identically zero at leading order, namely, $C_i^{(\Delta_Z)}(N_R) = C_i^{(\Delta_\gamma)}(N_R) = 0$. Moreover, $C_{i}^{(\Box)}(H^\pm) = C_{i}^{(\Delta_\gamma)}(H^\pm) = 0$ since in the 2HDM the charged Higgs contributes only to the $Z$ penguin diagram. 
Finally, $C_{i}^{(\Box)}(N_R, H^\pm)$ includes the contributions of the heavy neutrinos exchange mediated by a charged Higgs current and, as such, is peculiar of the model considered in this paper.  
The analytic expressions of the new physics contributions to the Wilson coefficients for the penguin diagrams are as follows. 
\begin{eqnarray}
&& \textrm{\bf{Penguins: contributions from the charged Higgs boson}}    \nn \\
&& C_9^{(\Delta_Z)}(H^\pm) = - \zeta_u^2 \left( \frac{-1 + 4 s_W^2}{8 s_W^2 }  \right) \frac{  \xHt \, x_t  ( \xHt -1  - \log \xHt )}{(\xHt - 1)^2},    \nn \\ 
&& C_{10}^{(\Delta_Z)}(H^\pm) = - \zeta_u^2 \left( \frac{1}{8 s_W^2 }  \right) \frac{  \xHt \, x_t  ( \xHt -1  - \log \xHt )}{(\xHt - 1)^2},      \nn \\
&& C_{9}^{(\Delta_\gamma)}(H^\pm) = \zeta_u^2 \frac{\xHt}{108 (\xHt-1)^4} \bigg[6 \left(3 \xHt^3-6 \xHt+4\right) \log \xHt  -  (\xHt-1) (\xHt (47 \xHt-79)+38)\bigg],      \nn \\
&& C_{10}^{(\Delta_\gamma)}(H^\pm) = 0.    \label{eq:peng-charged}     \\
\nn \\
&& \textrm{\bf{Penguins: contributions from the heavy neutrinos}}  \nn \\
&& C_i^{(\Delta_Z)}(N_R) = C_i^{(\Delta_\gamma)}(N_R) = 0 .
\end{eqnarray}

\noindent
The analytic expressions of the new physics contributions to the Wilson coefficients for the box diagrams are as follows:

\begin{eqnarray}
&& \textrm{\bf{Box: contributions from the charged Higgs boson}}  \nn \\
&& C_{9}^{(\Box)}(H^\pm) = C_{10}^{(\Box)}(H^\pm) = 0. \\
&& \nn \\
&& \textrm{\bf{Box: contributions from the heavy neutrinos}}  \nn \\
&& C_{9}^{(\Box)}(N_R) =  \sum_{i = 1}^{n_R}  | (U_{Lh})_{\ell i}  |^2 \frac{  x_t}{16 s_w^2 (x_{N_i}-1)
   (x_t-1)^2 (x_{N_i}-x_t)^2}   \bigg[(x_{N_i}-1) \left((x_t-1) (x_t-x_{N_i}) \right. \nn \\
 && \qquad \left. \times  (7
   x_t-4 x_{N_i})  \right. 
  - \left.  \left(4 x_{N_i}^2-x_{N_i} x_t (3
   x_t+8)+x_t^2 (6 x_t+1)\right) \log
   x_t \right) \nn \\
 && \qquad -(x_t-1)^2 \left(4 x_{N_i}^2-8 x_{N_i}
   x_t+x_t^2\right) \log x_{N_i} \bigg],  \nn \\
&& C_{10}^{(\Box)}(N_R) = - C_{9}^{(\Box)}(N_R). 
%
\end{eqnarray}
\begin{eqnarray}
&& \textrm{\bf{Box: contributions from the charged Higgs boson and the heavy neutrinos}}  \nn \\
&& C_{9}^{(\Box)}(N_R, H^\pm) =  \sum_{i = 1}^{n_R}  | (U_{Lh})_{\ell i}  |^2  \bigg\{
\frac{\zeta_u^2 \, \zeta_\nu^2 \, \xHt x_t}{16 s_W^2 (\xHt-1)^2 (x_{N_i}-1) (\xHt-x_{N_i})^2}  \nn \\
&& \qquad \times \bigg[-\xHt (\xHt-1)^2 \log x_{N_i} - (x_{N_i}-1) ((\xHt-1) (\xHt-x_{N_i}) \nn \\
&& \qquad +\xHt (-2
   \xHt+x_{N_i}+1) \log \xHt) \bigg] \nn \\
&& \qquad +   \frac{\zeta_u \, \zeta_\nu \, \xHt x_t}{8 s_W^2  (\xHt-1) (x_{N_i}-1) (x_t-1) (\xHt-x_{N_i}) (\xHt-x_t) (x_{N_i}-x_t)} \nn \\
&& \qquad \times
   \bigg[(x_{N_i}-1) (x_t-1) (4 \xHt-x_t) \log \xHt
   (x_{N_i}-x_t)  \nn \\
  && \qquad + (\xHt-1) ((x_t-1) (\xHt-x_t)
   (x_t-4 x_{N_i}) \log x_{N_i} +3 (x_{N_i}-1) x_t
   (\xHt-x_{N_i}) \log x_t  )\bigg]
   \bigg\}, \nn \\
&& C_{10}^{(\Box)}(N_R, H^\pm) = - C_{9}^{(\Box)}(N_R, H^\pm)  .
\end{eqnarray}

In the previous equations we used the following mass ratios
\bea
x_t =  \frac{M_t^2}{M_W^2}\,, \qquad x_H = \frac{M_t^2}{M^2_{H^\pm}} \,, \qquad x_{N_i} =  \frac{M_t^2}{m^2_{\nu_{hi}}}.
\eea
\begin{figure}[t]
\centering
\subfigure[]{\includegraphics[scale=1]{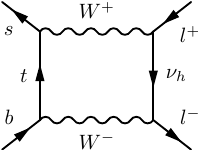}} \qquad \qquad 
\subfigure[]{\includegraphics[scale=1]{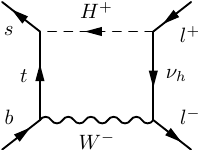}} \qquad \qquad 
\subfigure[]{\includegraphics[scale=1]{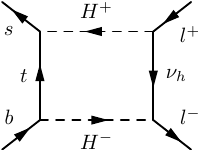}}
\caption{The heavy neutrino contributions, with and without the charged Higgs, to the one-loop box diagrams of the $b \to s \ell^+\ell^-$ transition.}
\label{fig:box}
\end{figure}

We now discuss some interesting features concerning the structure of these coefficients. First, one can see for the box diagram contribution in fig. \ref{fig:box} that the heavy neutrinos give $C_9 = - C_{10}$.  
One can also see that the SM box contribution with the light neutrinos, not shown above, is rescaled by $\sum_{i=1}^{3} |\tilde U_\textrm{PMNS}^{\ell i}|^2 \simeq 1 - \eta^2$, where $\ell = e, \mu$ and $\eta^2$ controlling the departure from unitarity of the PMNS matrix. (Since $\eta^2$ is expected to be small and no new physics enhancement factors are present in the SM diagram, we can safely neglect this correction.) 
Concerning the box diagrams, there is also a new contribution $C_{i}^{(\Box)}(N_R)$ with the heavy neutrinos and  virtual $W^\pm$s. 
This is proportional to $\sum_{i = 1}^{n_R}  | (U_{Lh})_{\ell i}  |^2$ and we do not expect, as confirmed by the numerical analysis, that this contribution can  provide large effects to  flavor observables. 
Indeed, the coupling of the heavy neutrinos to the leptons mediated by the $W^\pm$ boson is fixed by the gauge invariance and proportional to the $SU(2)$ weak gauge coupling. 
To allow for more freedom one has to rely on an extra charged degree of freedom with the simplest possibility being the charged scalar of a 2HDM extension. 
These contributions are encoded into $C_{i}^{(\Box)}(N_R, H^\pm)$.
The $Z_2$ symmetric scenarios of the 2HDM are among the simplest ones but barely produce significant effects in the $C_{9,10}$ Wilson coefficients.
This can be understood from Table~\ref{tab:2hdms}, because the corrections to $C_{9,10}$ would be proportional to $\zeta_u^2 = \zeta_\nu^2 = \cot^2 \beta$, independently from the specific realization, and thus relevant only for $\tan \beta < 1$, which is severely constrained by $b \to s \gamma$. 
The A2HDM allows us to disentangle $\zeta_u$ from $\zeta_\nu$, such that, while the former is still bound from $b \to s \gamma$, the latter can be varied freely, thus providing significant contributions to the Wilson coefficients in some region of the parameter space. 
We recall again that the alignment in the neutrino sector is not strictly required by the flavor physics but we, nevertheless, impose it by assuming that the same mechanism ensuring the proportionality between the Yukawa couplings is in place in both the quark and  lepton sectors.

\section{Results}

\begin{figure}[h]
\centering
\includegraphics[scale=0.6]{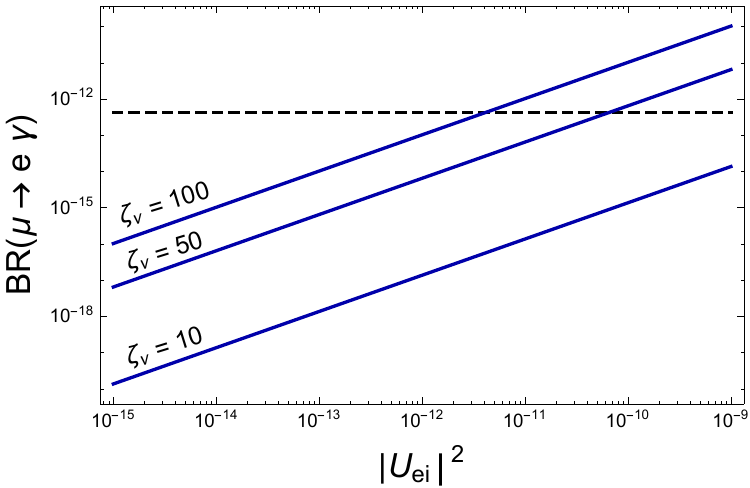}
\caption{$\textrm{BR}(\mu \to e \gamma)$ for $M_{H^\pm} = 700$ GeV, $m_{\nu_{h_i}} = 500$ GeV, $|(U_{Lh})_{\mu i}|^2 = 0.4 \times 10^{-3}$ and $\zeta_\ell = 0$. The dashed line corresponds to the MEG exclusion bound.}
\label{fig:BRmuega}
\end{figure}

Among the different lepton flavor-violating processes, $\mu \to e \gamma$ is the most constraining one; we use the MEG bound of $BR(\mu \rightarrow e \gamma) \leq 5.7 \times 10^{-13}$, from Ref. \cite{Adam:2013mnn}.
In Fig.~\ref{fig:BRmuega} we show the corresponding BR generated by a single heavy neutrino as a function of the squared mixing angle of the same heavy neutrino with the electron one. The other parameters have been fixed as $M_{H^\pm} = 700$ GeV, $m_{\nu_{h_i}} = 500$ GeV, the $\zeta_\ell = 0$, and $|(U_{Lh})_{\mu i}|^2 = 0.4 \times 10^{-3}$. The latter corresponds to the maximum allowed value from unitarity constraints; see Eq.~(\ref{eq:unitarity}).
Since in a model with a single Higgs doublet we would obtain $\textrm{BR}/U^4 \sim (6 - 7) \times 10^{-4}$ in the heavy neutrino mass range of $500 - 1000$ GeV, it is clear that the diagrams with the charged Higgs give the dominant and a very large contribution to the BR. 
Using the largest values for the mixing angles allowed by unitarity we obtain a suppression of the BR of $\sim 10^{-7}$ which cannot accommodate the strong bound on $\mu \to e \gamma$ from the MEG experiment. This suggests that, if a low-scale seesaw is embedded into a 2HDM framework, a given heavy neutrino (or, in a scenario with a large hierarchy between heavy neutrino states, the lightest one) may have a non-negligible mixing only with SM neutrinos of a given flavor eigenstate; otherwise, the charged Higgs boson would induce unacceptably large effects on lepton flavor-violating processes. The two realisations would be $(U_{Lh})_{\mu i} \neq 0, (U_{Lh})_{e i} \simeq 0$ or $(U_{Lh})_{\mu i} \simeq 0, (U_{Lh})_{e i} \neq 0$.
As we will see below, the first possibility can be also used to explain the deviation of $R_{K^*}$ and $R_K$ from the SM prediction. 
These two conditions can be achieved by suitably choosing the mass matrices $m_D$ and $m_R$.
Another possibility, allowing one to control the large effects in $\textrm{BR}(\mu \to e \gamma)$, which, anyway, we will not explore in this work, is to assume a $\zeta_\ell \neq 0$ and tune it against the $\zeta_\nu^2$ term in Eq.~(\ref{eq:GHpm}) to reduce the $\mathcal G_{H^\pm}$ form factor with respect to $\mathcal G_{W^\pm}$.

With the analytic expressions of the Wilson coefficients, we may turn to finding realistic benchmark scenarios.
We explore the parameter space spanned by the three parameters: $\zeta_\nu$, $m_{\nu_i}$ and $(U_{Lh})_{\mu, i}^2$ considering the impact of a single heavy neutrino, thus assuming, for the sake of simplicity, a hierarchy in the neutrino mass spectrum. The general case can be obtained straightforwardly and does not add much to the present discussion. 
In particular, one finds that if all the neutrino masses are almost degenerate, the combination $\sum_{i=1}^6 (U_{Lh})_{\mu, i}^2$ can be factored out from the Wilson coefficients and can be treated as an independent parameter leading to the same conclusions of the single neutrino case. 
This combination of squared mixing angles is bound from the nonunitarity test of the PMNS matrix to be less than $\sim 0.4 \times 10^{-3}$. 
Notice that we considered the scenario $(U_{Lh})_{\mu i} \neq 0, (U_{Lh})_{e i} \simeq 0$ in line with the previous discussion on lepton flavor-violating processes. 
The performed scan takes the ranges $-80 < \zeta_\nu < 80$, $10^{-5} < (U_{Lh})_{\mu, i}^2 <10^{-3},$ and $200 \textrm{ GeV} < m_{\nu_i} < 2000$ GeV.
The other parameters are chosen as $M_{H^\pm} = 550$ GeV, $\zeta_u =1$, and $\zeta_d = \zeta_\ell \simeq 0$, which ensure that the flavor constraints discussed above, namely, the ones which are not significantly affected by the presence of the heavy neutrinos, are all satisfied \cite{Enomoto:2015wbn}. Other choices are obviously acceptable but not considered here.

\begin{figure}[h]
\centering
\subfigure[]{\includegraphics[scale=0.45]{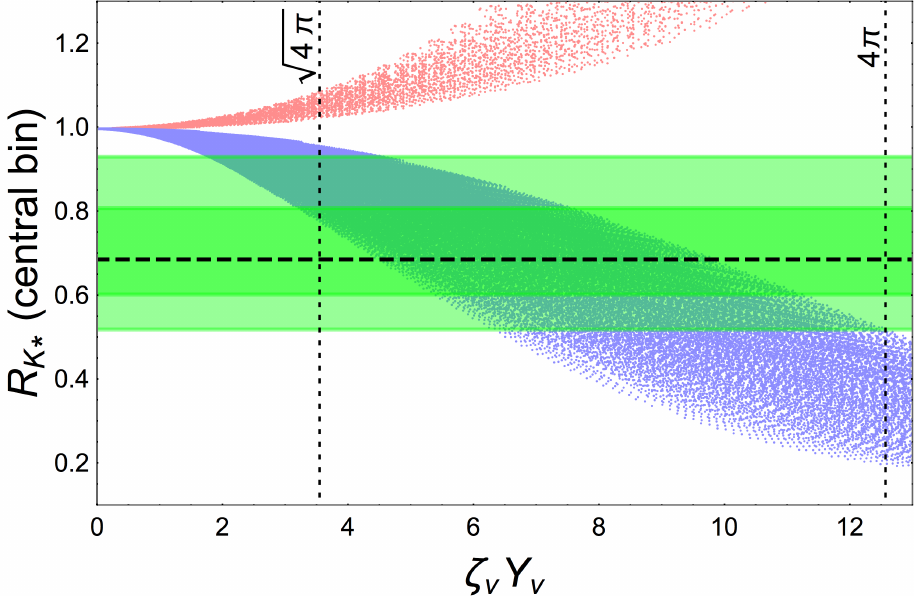}} \quad
\subfigure[]{\includegraphics[scale=0.45]{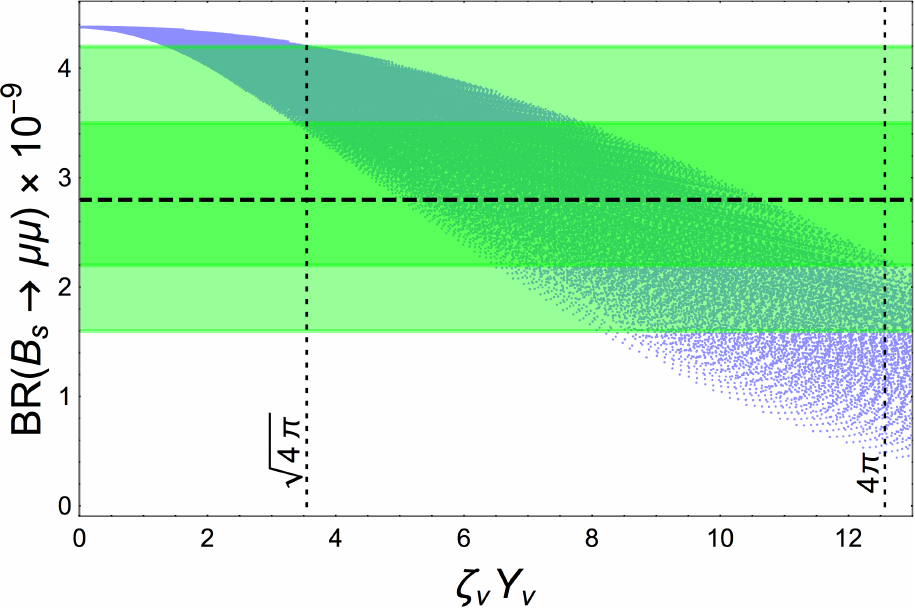}}
\caption{$R_{K^*}$ in the central bin and $\textrm{BR}(B_s \to \mu^+\mu^-)$ as a function of the combination of parameters that mostly controls the observable. 
The green bands are the 1$\sigma$ and 2$\sigma$ bands for the $R_{K^*}$ measurement. Blue (red) points correspond to the scenario in which the heavy neutrino has a non-negligible coupling only to the muon (electron) flavor eigenstates. }
\label{fig:RK_1}
\end{figure}

\begin{figure}[h]
\centering
\subfigure[]{\includegraphics[scale=0.45]{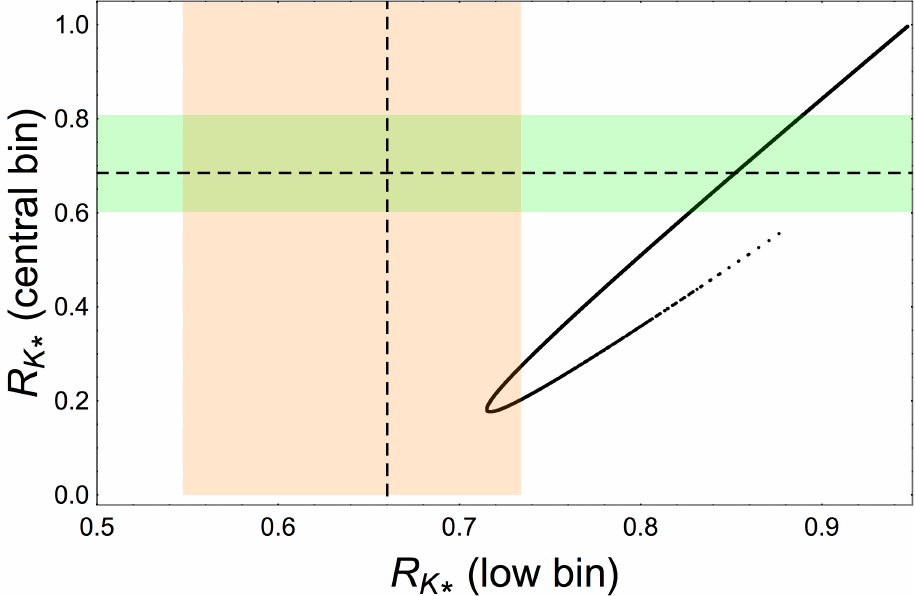}} \quad
\subfigure[]{\includegraphics[scale=0.45]{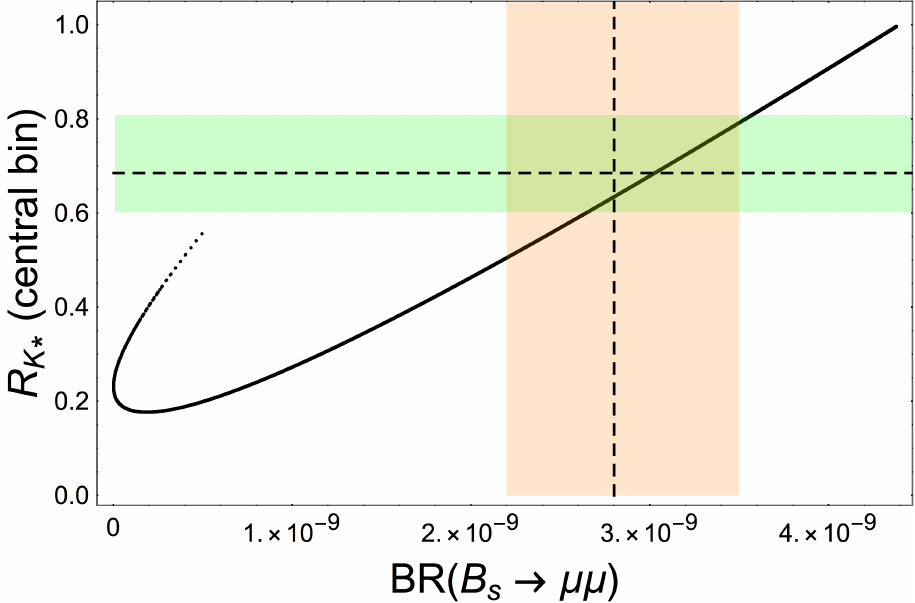}}
\caption{
(a) Correlation between $R_{K^*}$ in the central and  low bins.  
(b) Correlation between $R_{K^*}$ in the central bin and the predicted BR($B_s \to \mu^+ \mu^-$). Here only the 1$\sigma$ bounds have been considered in both figures.}
\label{fig:RK_2}
\end{figure}

The main result of the analysis is presented in Fig.~\ref{fig:RK_1}. 
The numerical values of the flavor observables $R_{K^*}$ and $R_K$ have been obtained from the evaluation of the Wilson coefficients computed above and from the \texttt{FLAVIO} package \cite{Straub:2018kue}. In the interesting region of the parameter space, the predictions for the two ratios in the A2HDM are the same, and, for the sake of simplicity, we will only discuss $R_{K^*}$.
The latter is presented as a function of the parameter $\zeta_\nu Y_\nu \equiv \sqrt{2} \, \zeta_\nu (U_{Lh})_{\mu i} \, (m_{\nu_{h_i}}/v)$ that mostly controls the $C_{9,10}$ Wilson coefficients.
In particular, the blue points correspond to the configuration in which the heavy neutrino couples to the SM muon sector and has negligible mixing with the first family. 
As anticipated above, this setup allows us to reproduce the measured reduction in the $R_{K^*}$ ratio, which is represented in the plot with a dashed horizontal line, together with 1$\sigma$ and 2$\sigma$ (green) bands. The red points, instead, are representative of the scenario in which the heavy neutrino has a mixing with the electrons. In this case, the predicted $R_{K^*}$ is above 1 and contradicts the LHCb observations.
The extent of the reduction is mainly controlled by the parameter $\zeta_\nu Y_\nu$. Interestingly, the same parameter defines the strength of the coupling among the charged Higgs,  heavy neutrino and  lepton and, as such, is subject to the perturbativity bound of Ref. \cite{Rose:2015fua}.  The upper limit for the coupling is usually extracted from naive scaling arguments but also depends on the loop functions of the involved processes. 

{
Besides perturbativity, the parameter $\zeta_\nu Y_\nu$ is restricted by the requirement that the total widths of the charged Higgs boson and  heavy neutrinos are not larger than their masses, since it controls, in particular, the partial decay widths $\Gamma (H ^\pm \rightarrow \nu_h \mu ^\pm)$ and $\Gamma ( \nu_h \rightarrow H^\pm \mu ^\mp )$. In the limit of a massless charged lepton, we find
\bea
\Gamma (H ^\pm \rightarrow \nu _\mu \mu ^\pm) = \frac{M_{H^+}}{16 \pi} (\zeta_\nu Y_\nu)^2 \left( 1 - \frac{m_{\nu_{h}}^2 }{M_{H^+}^2} \right)^2
\eea
as well as an analogous result for $\Gamma ( \nu_h \rightarrow H^\pm \mu ^\mp )$ with $m_{\nu_{h}} \leftrightarrow M_{H^+}$ and an extra factor of 2 to account for the  charged-conjugated final states. 
Ignoring the kinematic factor, we obtain the constraint $\zeta_\nu Y_\nu \lesssim 4 \sqrt{\pi}$ which lies between the two commonly adopted upper bounds that can be
used to estimate the region of tree-level perturbativity and that are shown in the plot.
One should keep in mind that, by adopting the most conservative bound, $\zeta_\nu Y_\nu \lesssim  \sqrt{4 \pi}$, it is possible to explain the present $R_{K^*}$ measurement and the $\textrm{BR}(B_s \to \mu^+\mu^- ) =(2.8 ^{+1.4} _{-1.2})\times 10^{-9}$ \cite{CMS:2014xfa} 
only at the $2 \sigma$ level.
}

{
While the limit $\zeta_{\nu} Y_\nu \rightarrow 0$ switches off the nonuniversal leptonic contributions, lepton universal effects beyond the SM predictions still survive. This is particularly clear in Fig.~\ref{fig:RK_1}(b). 
The latter shows that the BR$(B_s ^0 \rightarrow \mu \bar{\mu})$ is characterized by a correction of $\sim 20 \%$ with respect to the SM value for the chosen benchmark point.
These effects are intimately linked to the 2HDM nature of the BSM scenario discussed here and, in particular, to the presence of a charged Higgs state in the scalar spectrum.
The lepton universal contributions arise from the penguin diagrams and affect both $C_9$ and $C_{10}$ as detailed in Eq.~(\ref{eq:peng-charged}). 
The corrections to $C_{10}$ are larger ($\sim 10\%$) than those to $C_9$ in which case the $Z$-penguin and photon-penguin diagrams almost cancel each other. 
This feature breaks the $C_9 = -C_{10}$ effect induced by the heavy neutrinos. 
The impact of the charged Higgs boson in other flavor observables and the corresponding constraints have been discussed in Refs. \cite{Enomoto:2015wbn,Li:2014fea}. 
As stated above, we have ensured that the benchmark point considered complies with all the relevant bounds. 
The scalar charged current mainly affects neutral meson mixings ($B_q^0 - \bar B_q^0$ and $K^0 - \bar K^0$, with $\Delta M_s$ providing the strongest constraint),
leptonic decays of charged and neutral mesons (in particular, $B_s^0 \to \mu^+ \mu^-$, $B_d^0 \to \mu^+ \mu^-$, and $B \to \tau \nu$), and 
the radiative decay $\bar B \to X_s \gamma$. The 2HDM extension of the SM has also been employed to address the anomalous magnetic moment of the muon, $R(D)$ and $R(D^*)$; nevertheless, the explanation of these anomalies lies outside the motivation of this paper.
}

Figure~\ref{fig:RK_2}(a) displays the 1$\sigma$ bounds for both the $R_K ^{*}$ low and central bins. One can see that parameter configurations can satisfy either the low or central bins separately, but not simultaneously both. In Fig.~\ref{fig:RK_2}(b), we consider the effects on the BR$(B_s \rightarrow \mu^+ \mu^-)$. One can see that for several points both predictions are simultaneously compatible with the experimental measurements.

\section{Conclusions}
The LHCb experiment at CERN has recently reported the existence of some anomalies in their data, with respect to the predictions of the SM. Specifically, 
the measured values of the observables $R_{K}= {\rm BR}(B^+\to K^+ \mu^+\mu^-)/{\rm BR}(B^+\to K^+ e^+ e^-)$
and $R_{K^*}= {\rm BR}(B^0\to K^{*0} \mu^+\mu^-)/{\rm BR}(B^0\to K^{*0} e^+ e^-)$ revealed a $\approx2.5 \sigma$ deviation when compared to the SM rates, which are essentially 1. In addition, the discrepancies occur in two dilepton invariance mass bins. Therefore, these results must be taken as a serious hint of possible BSM physics.

Herein, we have considered the possibility of  explaining such $R_K$ and $R_{K ^*}$ anomalies in an A2HDM, wherein  an alignment is present between the Yukawa couplings generated by the two Higgs doublet fields, combined with a low-scale seesaw mechanism generating light neutrino masses and mixings in compliance with current experimental measurements. Such a scenario allows for  significant nonuniversal leptonic  contributions, through box diagrams mediated by $H^\pm$ and $\nu_R$ states, which in turn alter the yield of the partonic decay $b \to s \ell^+ \ell^-$ entering the definition of 
 both the $R_K$ and $R_{K^*}$ observables. To render our explanation phenomenologically viable, we have made sure to comply with both theoretical (chiefly, the unitarity bounds stemming from the neutrino mixing matrix) and experimental (the strongest being those due to lepton flavor-violating processes and mesonic decay channels) constraints. In fact, the masses required for the charged Higgs and  right-handed neutrino states entering the above transition are also well beyond their current direct limits. Furthermore, to make clear that our explanation of the  $R_K$ and $R_{K ^*}$ anomalies is not particularly {\sl ad hoc}, we have left the actual low-scale dynamics onsetting the seesaw mechanism undetermined, by illustrating that this could be realized through different scenarios, e.g., the so-called inverse and linear seesaw cases. Therefore, our setup captures a variety of  light neutrino masses and mixings that can be tuned to further experimental observation in the neutrino sector while leaving predictions in the $B$ one unchanged. Finally, we have  correlated our predictions for $R_K$ and $R_{K ^*}$ in both dilepton invariant mass bins to those for the highly constraining observable BR$(B_s\to \mu^+\mu^-)$, showing that simultaneous solutions to both sets of measurements can be found in the envisioned A2HDM plus low-scale seesaw scenario.

\section*{Acknowledgements}
The work of L.D.R. and S.M. is supported in part by the NExT Institute. S.M. also acknowledges partial
financial contributions from the STFC Consolidated Grant No. ST/L000296/1. S.J.D.K. and S.K. have received support under the H2020-MSCA InvisiblesPlus
(RISE) Grant No. 690575 and Elusives (ITN) Grant No. 674896. All authors acknowledge support under the H2020-MSCA  NonMinimalHiggs (RISE) Grant No. 645722. L.D.R. thanks Fady Bishara, Marco Nardecchia and Olcyr Sumensari for very fruitful discussions.

\bibliographystyle{JHEP}
\providecommand{\href}[2]{#2}\begingroup\raggedright\endgroup
\end{document}